\documentclass[fleqn,twoside]{article}
\usepackage{espcrc2}
\usepackage{graphicx}
\usepackage[figuresright]{rotating}

\newcommand{\AmS}{{\protect\the\textfont2
  A\kern-.1667em\lower.5ex\hbox{M}\kern-.125emS}}
\newcommand{\nc}{\newcommand}
\nc{\ttbar}{t\bar{t}}         \nc{\bbbar}{b\bar{b}}
\nc{\tanb}{\tan \beta}        \nc{\twbdec}{t\to W^+ b}
\nc{\tbwbdec}{\bar{t}\to W^- \bar{b}}
\nc{\epem}{e^+e^-}            \nc{\eett}{\epem \to \ttbar}
\nc{\sigeett}{\sigma_{e\bar{e}\to\ttbar}}
\nc{\wpwm}{W^+W^-}            \nc{\tbar}{\bar{t}}
\nc{\bbar}{\bar{b}}           \nc{\wpp}{W^+}
\nc{\mt}{m_t}    \nc{\mts}{m_t^2}   \nc{\mw}{m_W}    \nc{\mws}{m_W^2}
\nc{\mz}{m_Z}    \nc{\mzs}{m_Z^2}
\nc{\ttbardec}{\ttbar \to W^+W^-\bbbar}
\nc{\wwbb}{W^+W^-\bbbar}      \nc{\sm}{SM}
\nc{\cw}{\cos\theta_W}        \nc{\sw}{\sin\theta_W}
\nc{\sws}{\sin^2\theta_W}     \nc{\sig}{\sigma_{tot}}
\nc{\lp}{\ell^+}              \nc{\lm}{\ell^-}
\nc{\epsl}{\epsilon_L}        \nc{\cp}{C\!P}
\nc{\pp}{\gamma \gamma}
\nc{\pptt}{\pp \to \ttbar}
\nc{\ocal}{{\cal O}}
\nc{\lspace}{\;\;\;\;\;\;\;\;\;\;}  \nc{\llspace}{\lspace \lspace}
\nc{\non}{\nonumber}
\nc{\tb}{\stackrel{{\scriptscriptstyle (-)}}{t}}
\nc{\bb}{\stackrel{{\scriptscriptstyle (-)}}{b}}
\nc{\fb}{\stackrel{{\scriptscriptstyle (-)}}{f}}
\nc{\beq}{\begin{equation}}   \nc{\eeq}{\end{equation}}
\nc{\bea}{\begin{eqnarray}}   \nc{\eea}{\end{eqnarray}}
\nc{\baa}{\begin{array}}      \nc{\eaa}{\end{array}}
\nc{\bit}{\begin{itemize}}    \nc{\eit}{\end{itemize}}
\nc{\ben}{\begin{enumerate}}  \nc{\een}{\end{enumerate}}
\nc{\bce}{\begin{center}}     \nc{\ece}{\end{center}}
 \nc{\re}{\hbox {Re}} 
\title{Effects of Top-Quark Anomalous Decay Couplings 
at $\gamma \gamma$ Colliders\thanks{This is based on collaboration with
B. Grzadkowski, Z. Hioki, and J. Wudka.}}

\author{
{\sc Kazumasa}  OHKUMA\thanks{
E-mail address: ohkuma@radix.h.kobe-u.ac.jp}\vspace*{0.1cm}\\
{Graduate School of Science and
Technology, Kobe University, Nada, Kobe 657-8501, JAPAN}%
}
\begin{document}
\begin{abstract} 
Most general $tbW$ couplings were investigated
in the process $\gamma \gamma \rightarrow t\bar{t} 
\rightarrow \ell^{\pm} X$ for unpolarized photon beams.
The double angular and energy distribution of the lepton was calculated 
and an optimal-observable analysis on it was carried out 
for the SM $t\bar{t}$ production mechanism.
It was also shown that the leptonic angular distribution
is insensitive to non-standard $tbW$ vertex. 
That means that observation of non-standard effects indicates 
existence of some new physics in the production part.
\vspace{1pc}
\vspace{-1cm}
\end{abstract}
\maketitle
\setcounter{footnote}{0}
\section{INTRODUCTION}
The standard model (SM) is extremely successful in particle-physics
phenomenology. 
However, since there are still several unsolved problems  in the SM,
it is  believed that the new physics beyond the SM is necessary to solve
them.  Therefore, it is very important to search for signals of new physics
beyond the SM at forthcoming collider experiments.

Top-quark interactions could prove to be a reach source of information on
the new physics. 
Since the mass of the top quark is much larger than the masses of the
other quarks and leptons, and couplings of the top quark are still not
strongly constrained,  
it is very interesting to investigate contribution from possible new physics
in the top-quark sector.
In addition, the top quark has some advantageous properties for searching
new physics beyond the SM~\cite{rept}. 
Therefore, in this work, we will discuss effects from possible 
anomalous $tbW$ couplings in the process
$\gamma \gamma \rightarrow t\bar{t}\rightarrow \ell^{\pm}X$, which
will be observed at future photon-photon colliders,
by performing a model-independent form-factor analysis~\footnote{
Non-standard effects in top-quark productions at $\gamma\gamma$ collider have
been studied in ref.\cite{ch}. 
}.
%
\section{FRAMEWORK}
We will adopt the Kawasaki-Shirafuji-Tsai formula~{\cite{kst}},
assuming that  both decaying top quarks
and $W$ bosons are  on-shell particles.
In this formula, production and decay part
are factorized as follows:
\begin{eqnarray}
&&\hspace*{-0.7cm}\it
\frac{d \sigma}{d {\mbox{\boldmath{$p$}}}_{\ell}}(\gamma \gamma \to \ell^+ X)
=\: {\rm 4} \int d \Omega_t \frac{d \sigma}{d \Omega_t}(n_t,{\rm 0}) \non \\
&&\it \hspace*{2.5cm}
\times\frac{1}{\Gamma_t}\frac{d \Gamma}{d {\mbox{\boldmath{$p$}}}_{\ell}}
(t\to b \ell^+ \nu) ,\label{eq1}
\end{eqnarray}
where ${\it \Gamma}$ and ${\it \Gamma_t}$ are the leptonic and total widths of 
the top quark, respectively,  i.e., $\it \Gamma=B_{\ell}\Gamma_t$, 
using $\it B_{\ell}$ as the branching ratio for $t\to b\ell^+ \nu$
($\simeq$ 0.22).
$\it {d \Gamma}/{d {\mbox{\boldmath{$p$}}}_{\ell}}$
means the differential decay rate for the top quark.
$\it d \sigma (n_t, {\rm 0})/d \Omega_t$ is  obtained from
$d \sigma (s_t, s_{\bar{t}})/ d \Omega_t$ which is  
the angular distribution of $t\bar{t}$ with spin $s_{t}$ and
$s_{\bar{t}}$
by replacing
$\it  s_{t}^{\mu} \to n_{t}^{\mu}=
\left(
g^{\mu \nu}-\frac{p_t^\mu p_t^\nu}{m_t^{\rm 2}} 
\right)
\frac{m_t}{({p_t p_\ell})} p_{\ell\nu}
$
and setting $s_{\bar{t}}\to 0$~\cite{aandz,bandz2}.

Since we will perform a model-independent analysis
we should use the most general couplings of $tt\gamma$
and $tbW$.
However, since 
top-quark pair is produced via t- and  u-channel processes 
in $\gamma \gamma$ collisions differently from the
case of $e^+ e^-$ collisions in which top-quark pair is produced 
via s-channel one, there are two $tt\gamma$ couplings,
and each $ tt\gamma$ coupling includes one virtual top
quark.
Thus, without some assumptions or choice of specific models,
it is not possible to estimate momentum 
dependence of form factors of $tt\gamma$ coupling,
and consequently 
we cannot perform the $d{\mit\Omega}_t$ integration in (\ref{eq1}).

On the other hand, in top-quark decay processes,
since top quarks, bottom quarks  and $W$ bosons can
be treated as on-shell particles,  
we can fix  momentum dependence of form factors and
parameterize non-standard effects in terms of constant parameters.
For these reasons, in this work,  we will focus on the 
possibilities of new physics  in the decay part  
assuming the SM production of the top quark.

We will adopt the following 
most general form of the $tbW$ vertex suitable for 
the $t\rightarrow W^+ b$  
decays~\footnote{Of course, though the vertex for $\bar{t}\to W^- \bar{b}$
decay would  be also written down, it was omitted.}: 
\begin{eqnarray}
&&\!\!{\mit\Gamma}^{\mu}_{Wtb}=-{g\over\sqrt{2}}\:
\bar{u}(p_b)\biggl[\,\gamma^{\mu}(f_1^L P_L +f_1^R P_R) \biggr.\non \\
&&\phantom{\!\!{\mit\Gamma}^{\mu}_{Wtb}}
\biggl. 
-{{i\sigma^{\mu\nu}k_{\nu}}\over M_W}
(f_2^L P_L +f_2^R P_R)\,\biggr]u(p_t),\ \ \ \ \ \ \label{ffdef}
%
\end{eqnarray}
where $P_{L/R}=(1\mp \gamma_5)$ and $k$ is 
the momentum of $W$ boson.
$f_{1,2}^{L/R}$ parameterize non-standard decay effects.
Because $W$ boson is on-shell in the narrow-width approximation,
the two additional form factors do not contribute. 
In the SM, $f_1^L=1$ and the other form factors 
vanish.

Eq.(\ref{ffdef}) allows us to
perform global analysis of the non-standard effects
in the top-quark  decay if the decay is described by the sequential 
chain: $t \to W^+ b \to b\ell^+\nu_\ell$.
\section{FINAL LEPTON DISTRIBUTION}
In the following calculation,
we neglected masses of final leptons,
bottom quarks and quadratic terms of non-standard 
form factors.
\subsection{Angular and Energy Distribution}

Adopting the SM production of $t\bar{t}$ and the
form factors of the top-quark decay in eq. (\ref{ffdef}),
we can calculate the angular and energy distribution of 
final leptons as 
\begin{eqnarray*}
&&\hspace*{-0.5cm}
\frac{d\sigma}{dx d\cos\theta_{\ell}}(\gamma\gamma\to {\ell}^+ X)
=\frac{3(eQ_t)^4\beta}{64\pi^2 s}\\
&&\hspace*{-0.5cm}
\phantom{aaaa}\times \Bigl[\:f_1(x, \cos\theta_{\ell})
+{\rm Re}(f_2^R) f_2(x, \cos\theta_{\ell}) \:\Bigr],
\end{eqnarray*}
where
\begin{eqnarray*}
&&\hspace*{-0.5cm}
f_1(x, \cos\theta_{\ell})
=x \Bigl[\:-2\pi{\cal G}_{00} \Bigl.\\
&&\hspace*{1.3cm}
        +\Bigl(1+\frac{4m_t^2}{s}-\frac{8m_t^4}{s^2}\Bigr)
{\cal G}_{01}
 \Bigl.-\frac{8m_t^4}{s^2}{\cal G}_{02}\:\Bigr], \\
&&\hspace*{-0.5cm}
f_2(x, \cos\theta_{\ell})
=2\sqrt{r} x \Bigl[\:-2\pi{\cal G}_{10} \Bigl.\\
&&\hspace*{1.0cm}
        +\Bigl(1+\frac{4m_t^2}{s}-\frac{8m_t^4}{s^2}\Bigr){\cal G}_{11}
        -\frac{8m_t^4}{s^2}{\cal G}_{12}\:\Bigr],
\end{eqnarray*}
with
\begin{eqnarray*}
&&\hspace*{-0.6cm}
{\cal G}_{00}(x, \theta_{\ell})\equiv
C\int^{c_+}_{c_-} d\cos\theta_t \:\omega,\\
&&\hspace*{-0.6cm}
{\cal G}_{01}(x, \theta_{\ell})
\equiv C\int^{c_+}_{c_-} d\cos\theta_t \:\omega(I_+ +I_-),\\
&&\hspace*{-0.6cm}
{\cal G}_{02}(x, \theta_{\ell})\equiv
C\int^{c_+}_{c_-} d\cos\theta_t \:\omega(J_+ +J_-),\\
&&\hspace*{-0.6cm}
{\cal G}_{10}(x, \theta_{\ell})\equiv
C\int^{c_+}_{c_-} d\cos\theta_t \:\omega
\Bigl(\frac{1}{1-\omega}-\frac{3}{1+2r} \Bigr),\\
&&\hspace*{-0.6cm}
{\cal G}_{11}(x, \theta_{\ell})\equiv
C\int^{c_+}_{c_-} d\cos\theta_t \:\omega
\Bigl(\frac{1}{1-\omega}-\frac{3}{1+2r} \Bigr)\\
&&\phantom{aaaaaa}\times(I_+ +I_-), \\
&&\hspace*{-0.6cm}
{\cal G}_{12}(x, \theta_{\ell})\equiv
C\int^{c_+}_{c_-} d\cos\theta_t \:\omega
\Bigl(\frac{1}{1-\omega}-\frac{3}{1+2r} \Bigr)\\
&&\hspace*{-0.6cm}
\phantom{aaaaaa}\times(J_+ +J_-),\\
&&\hspace*{-0.7cm}
I_{\pm} 
\equiv \int_0^{2\pi}dx\frac1{A_{\pm}\cos x +B_{\pm}}
=2\pi/\sqrt{B_{\pm}^2-A_{\pm}^2},\\
&&\hspace*{-0.7cm}
J_{\pm} 
\equiv \int_0^{2\pi}dx\frac1{(A_{\pm}\cos x +B_{\pm})^2}\\
&&\hspace*{-0.1cm}
=2\pi B_{\pm}/\sqrt{(B_{\pm}^2-A_{\pm}^2)^3}\\
&& \hspace*{-0.7cm}
A_{\pm}=\pm \beta\sin\theta_t \sin\theta_{\ell},~   
B_{\pm}=1\pm\beta\cos\theta_t \cos\theta_{\ell},\\
&& \hspace*{-0.7cm}
r \equiv \left({M_W}/{m_t}\right)^2,~
 W \equiv (1-r)^2 (1+2r),\\ 
&& \hspace*{-0.7cm}
\omega \equiv {(p_t - p_{\ell})^2}/{m_t^2},~
C \equiv {3B_{\ell}(1+\beta)^2 s}/({m_t^2 W}).
\end{eqnarray*}
In the above integration, $c_{\pm}$ express the kinematical 
upper/lower bounds of $\cos\theta_t$ (see ref.~\cite{bandz2}), 
$x$ is the rescaled energy of the final lepton 
defined in terms of  its CM-frame energy $E_{\ell}$  and the 
top-quark velocity $\beta(\equiv \sqrt{1-4m_t^2/s})$ as
$
x \equiv \frac{2 E_{\ell}}{m_t}\sqrt{{(1-\beta)}/{(1+\beta)}},
$
and $\theta_{\ell}$ is the angle between the initial beam
direction and the final-lepton momentum.
This result shows that
the angular and energy distribution enables us to investigate 
non-standard decay via ${\rm Re}(f_2^R)$  exclusively.

Performing  an optimal-observable analysis~\cite{ooa} for 
$t$ decay,
we have determined the following size of anomalous coupling 
that would guarantee a signal at $1\sigma$ level:
$
\delta({\rm Re}(f_2^R))=
8.58\times 10^{-3}
$
with  integrated luminosity $L=500~{\rm fb}^{-1}$ and
one-lepton-detection efficiency $\epsilon=0.6$.
Of course we can carry out  an analogous calculation for 
$\bar{t}$ decay, and get the same result.
%
\subsection{Angular Distribution}

Performing further integration over $x$ for
$r(1-\beta)/(1+\beta)\leq x \leq 1$,
we obtain the angular distribution of the final lepton:\vspace{-0.1cm}
\begin{eqnarray*}
&&\hspace*{-0.5cm}
\frac{d\sigma}{d\cos\theta_{\ell}}
=\frac{(eQ_t)^4\beta }{128\pi^2 s}\:
(r-1)^2 \:(1+2r)\:(\beta -1)^2\\
&&\hspace*{0.8cm}
\times \:C\int^{c_+}_{c_-}
\frac{d \cos\theta_t}{(1-\beta \cos\theta_t )^{2}}\:
\left[\:-2\pi\phantom{\frac{A}{B}}\right.\\
&&\hspace*{0.5cm}\left. 
+\:\left(1+\frac{4m_t^2}{s}-\frac{8m_t^4}{s^2}
\right) 
(I_++I_-)\right.\\
&&\hspace*{2.0cm}
\left.-\frac{8m_t^4}{s^2}\:(J_++J_-)\:
\right].
\end{eqnarray*}
Surprisingly, we found that $f_2^R$ terms
vanish 
after $x$ integration. This is one typical example of the
general decoupling theorem found in \cite{bandz}.
For $d \sigma(\gamma \gamma \rightarrow \ell^- X)/d \cos \theta_{\ell}$,
same calculations lead us to an analogous result.
Thus, in this distribution, we should not observe any signal of
non-standard top-quark decay.

This means: if non-standard effects are observed in this
distribution, that is a signal of some physics
beyond our assumptions.
Let us recall what assumptions we have adopted 
in this work;
\begin{enumerate}
\item  \vspace*{-0.1cm}
       top-quark decay is described by sequential decays
        $t\to W^+b \to b\ell^+ \nu_{\ell}$,
\item \vspace*{-0.3cm}
        narrow-width approximation is applicable for $t$ and $W$,
\item \vspace*{-0.3cm}
        masses of $b$ quarks and final leptons, and  
        quadratic terms of non-standard form factors were
        neglected,
\item\vspace*{-0.3cm}
       non-standard effects do not exist in the production.
\end{enumerate}\vspace*{-0.3cm}
The assumptions 1$\sim$3 are well justified in this process, therefore
observation of non-standard effects in the angular distribution
would be a signal of physics beyond the assumption 4, i.e.
new physics 
in the production part.
\section{SUMMARY}
In the process
$\gamma \gamma \rightarrow t \bar{t} \rightarrow \ell^{\pm}X$
for unpolarized  photon beams\footnote{The analysis that takes into account
the beam polarization is under way\cite{ghow}.},
the angular and energy distribution,  and the angular distribution
of the secondary lepton 
were calculated, assuming the standard top-quark production and
the most general couplings for the decay.
In addition, the masses of bottom quarks and final leptons have 
been neglected and only linear terms of non-standard form factors 
have been kept. 
To calculate distributions of final leptons,
Kawasaki-Shirafuji-Tsai formula was adopted with narrow-width
approximation for the top quark and $W$ boson. 
                                        
The double angular and energy distribution is sensitive to the
non-standard contribution from the top-quark decay vertex.
Thus this observable seems to be a useful tool to measure
deviations from the standard top-quark decay. 
For the angular distribution of the final leptons,
non-standard effects of $tbW$ couplings 
completely vanish after integration of the energy dependence.
Therefore, an observation of any non-standard signal in the angular 
distribution may be an indication of some 
new physics in the production mechanism.


\end{document}